\documentclass[pre,twocolumn,showpacs] {revtex4-1}
\usepackage{graphicx}

\begin{document}
\title{Recognition of stable distribution with L\'evy index alpha close to 2}

\author{Krzysztof Burnecki}
\email{krzysztof.burnecki@pwr.wroc.pl}
\author{Agnieszka Wy{\l}oma{\'n}ska}
\email{agnieszka.wylomanska@pwr.wroc.pl}
\affiliation{Hugo Steinhaus Center, Institute of Mathematics and Computer Science,
Wroclaw University of Technology, Wroclaw 50-370, Poland 
}
\author{Aleksei Beletskii}
\email{beletskii@kipt.kharkov.ua}
\affiliation{Institute for Plasma Physics, National Science Center ``Kharkov Institute of Physics and Technology'', Kharkov 61108, Ukraine}
\author{Vsevolod Gonchar}
\email{vsevolod.gonchar@gmail.com}
\affiliation{Akhiezer Institute for Theoretical Physics, National Science Center ``Kharkov Institute of Physics and Technology'', Kharkov 61108, Ukraine}
\author{Aleksei Chechkin}
\email{achechkin@kipt.kharkov.ua}
\affiliation{Akhiezer Institute for Theoretical Physics, National Science Center ``Kharkov Institute of Physics and Technology'', Kharkov 61108, Ukraine}
\affiliation{Institute of Physics and Astronomy, University of Potsdam, 14476 Potsdam-Golm, Germany}
%\affiliation{Hugo Steinhaus Center, Institute of Mathematics and Computer Science,
%Wroclaw University of Technology,
%Wyspianskiego 27, 50-370 Wroclaw, Poland}
\begin{abstract} We address the problem of recognizing alpha-stable L\'evy distribution 
with L\'evy index close to 2 from experimental
data. We are interested in the case when the sample size of available data is not large, thus the power law 
asymptotics of the distribution is not clearly detectable, and the shape of empirical probability 
density function is close to a Gaussian. We propose a testing procedure combining a simple visual
test based on empirical fourth moment with the Anderson-Darling  and Jarque-Bera statistical tests and we check the efficiency of the method on simulated data. Furthermore,
we apply our method to the analysis of turbulent plasma density and potential fluctuations
measured in the stellarator type fusion device and demonstrate 
that the phenomenon of L-H transition
occurring in this device is accompanied by the transition from L\'evy to Gaussian fluctuation statistics.      
\end{abstract}
\maketitle
\section{Introduction}
Stable distributions, also called alpha-stable, or L\'evy stable \cite{Levy}, are ubiquitous in nature due to Generalized Central Limit Theorem.
It says that the stable distributions, like the Gaussian one, attract distributions of sums of 
independent identically distributed random variables \cite{Gnedenko}. Due to this reason, 
L\'evy stable distributions naturally appear when evolution of a system or result of an experiment 
are determined by a sum of many random factors. An important parameter characterizing stable distribution
is its index of stability, or the L\'evy index $\alpha$, $0 < \alpha \leq 2$. If $\alpha$ is strictly less than 2, the probability density function (PDF) of the stable law
exhibits slowly decaying power law asymptotic behavior of the form $|x|^{-1-\alpha}$ (``heavy tail''),
hence, the variance diverges. Due to this reason, the stable PDFs 
appear naturally in the description of random processes with large outliers, far from
equilibrium. The limiting case $\alpha = 2$ gives Gaussian distribution possessing fast decaying tails
and finite variance. In general, each stable distribution is characterized by four parameters:
L\'evy index $\alpha$, skewness parameter $\beta$, $-1 \leq \beta \leq 1$, scale parameter $\gamma$,
and location parameter $\delta$. For the Gaussian distribution $\beta$ is irrelevant, and it is
characterized by standard deviation and mean. The comprehensive theory of alpha-stable L\'evy distributions 
is presented in monographs \cite{Zolotarev, Samorodnitsky, Weron}.

Some measurable quantities  obey L\'evy statistics exactly, such as, e.g., gravitational field of masses
distributed randomly in space, hitting times for one-dimensional Brownian motion, and points of arrival in
two-dimensional one \cite{Feller}; interesting examples can be also found in 
\cite{Zolotarev, Uchaikin}. L\'evy statistics may also appear asymptotically due to Generalized 
Central Limit Theorem, like, e.g., in non-Brownian random walks with jumps and/or waiting times
obeying heavy-tailed distributions, see the reviews \cite{Metzler_1, Chechkin, Metzler_2}; 
we mention here such illuminating
examples as circulation of dollar bills \cite{Brockmann}, L\'evy flights for light in fractal
medium called L\'evy glass \cite{Barthelemy}, and L\'evy-like behavior of the marine vertebrates in response to patchy distribution of food resources \cite{Sims}. Stably distributed random noises are observed 
in such diverse applications as plasma physics (density and electric field fluctuations \cite{Gonchar, Kaw}), stochastic climate dynamics \cite{Ditlevsen}, physiology (heartbeats \cite{Peng}), electrical
engineering \cite{Nikias}, biology \cite{burwer10}, and economics \cite{Mantegna}. These and a lot of
other observations of stably (or stable-like) distributed quantities require reliable methods 
of random data analysis which allow to detect stable distributions (or distributions which belong to their
domains of attraction) and estimate their parameters. 

There are at least three procedures for estimating L\'evy stable law parameters: (i) the maximum likelihood 
method based on numerical approximation of the L\'evy stable likelihood function \cite{DuMouchel and Nolan};
(ii) the quantile method using tabulated quantiles of L\'evy stable laws \cite{McCulloch};
and (iii) the method using regression on the sample characteristic function \cite{Koutrouvelis and Kogon}.
All presented methods work well assuming that the sample under 
consideration is indeed L\'evy stable. The regression method is both fast and accurate, 
thus we use it in our analysis below. However, if the data come from a different distribution,
these procedures may mislead more than the Hill \cite{Hill} and direct log-log scale 
tail estimation method \cite{Rafal_1,janczura}, which focus only on the tail behavior without 
assuming parametric form for the whole distribution function.

The present paper, motivated by the analysis of plasma data (see Section V below), addresses
the issue of recognizing stable distributions with the L\'evy index close to 2. 
In this case, if the data sample is not large enough, the shape of empirical PDF is close to a Gaussian,
and both log-log scale analysis and Hill estimator give overestimated value of the
L\'evy index for the number of observation less than $10^6$  \cite{Rafal_2}, see also examples
in Figs. 3--7 of Ref. \cite{Rafal_1}. In applications,
especially in physics and biology, the number of observations is often less. In particular,
we deal with plasma data containing a few thousand of data points. Thus,
one has to be cautious in that domain of the L\'evy indices. Here we propose a certain
testing algorithm which, to our belief, could be helpful for the data analysis in that case. 
It combines ``visual inspection'' of L\'evy-type behavior with a few statistical tests, namely, 
the Anderson-Darling  and Jarque-Bera. Being direct and relatively simple, 
they, however, are rarely used in physical and biological applications, to our knowledge.
We demonstrate effectiveness of the proposed testing procedure by using simulated data
with known L\'evy stable distribution. As an application, we analyze the PDFs of plasma 
fluctuations measured in the fusion plasma device and detect an interesting change of 
plasma fluctuation statistics from L\'evy stable to a Gaussian one in the so-called L-H 
transition which occurs in the plasma device during operation. 

The structure of the paper is as follows. In Sec. \ref{simul} we first plot empirical PDFs for simulated
data with the L\'evy index close to $2$ and 5000 observations, and make sure that the difference 
from the Gaussian distribution is almost not visible. Then we plot the cumulative fourth
moment of the simulated data as a function of observation number and demonstrate that 
this simple visual test may help to distinguish between the L\'evy stable and Gaussian
PDFs. In Sec. \ref{schematt} we outline the algorithm used to distinguish between the L\'evy
stable and Gaussian distributions and demonstrate how this algorithm works for simulated
data. This is accompanied by a short sketch of the statistical tests which we employ for
the analysis. In Sec. \ref{appl} we apply the testing algorithm for the analysis of the experimental
data from plasma physics. Finally, in Sec. \ref{disc} we add a few remarks on the two
issues: the proposed testing procedure and change of statistics observed in Sec. \ref{appl}.

\section{Simulations}\label{simul}
\begin{figure}[htb]
\centering \resizebox{0.46\textwidth}{!}{\includegraphics{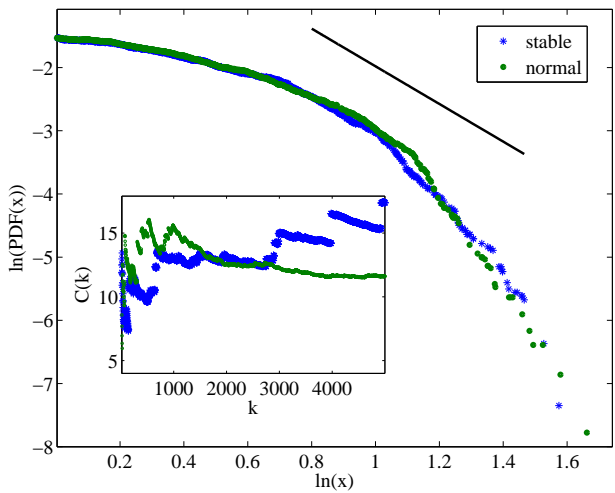}}
\caption{(Color online) The two empirical PDFs calculated for simulated samples from  L\'evy stable distribution with 
parameters $\alpha=1.98$, $\beta=0$, $\sigma=1$ and $\mu=0$, and from Gaussian distribution with mean 
equal zero and 
variance equal $2$, are shown in the log-log scale. Number of observations in both samples is $5000$. 
The straight line shows
the asymptotic slope of the L\'evy stable PDF with $\alpha=1.98$. Inset: Empirical cumulative fourth moment 
for the samples considered.}\label{fig11}
\end{figure}

In order to demonstrate behavior of the L\'evy stable distribution with L\'evy index close to $2$ we first turn to the analysis of simulated samples. The procedures of simulating stable random variables one can find in \cite{Weron,Rafal_2}.
In Fig. \ref{fig11} we  present in the log-log scale the right tail of the empirical PDF of the symmetric stable distribution with parameters $\alpha=1.98$, $\beta=0$, $\sigma=1$ and $\mu=0$ together with the empirical PDF of the Gaussian distribution with mean equal to zero and variance equal to 2. Let us mention, the Gaussian distribution with zero mean and variance equal to $2$ corresponds to the L\'evy stable one with $\alpha=2$, $\sigma=1$ and $\mu=0$ (the $\beta$ parameter is irrelevant in this case).
In both examined samples number of observations is set to $5000$.  Moreover, in Fig. \ref{fig11} we also 
show the straight line with the slope parameter $-2.98$ corresponding to asymptotics of the L\'evy stable PDF with $\alpha=1.98$.   As we observe, the difference between the two empirical PDFs is hardly visible.
Therefore, we propose to use a simple visual test to distinguish between the distributions. Namely, we calculate the empirical cumulative fourth moment (ECFM) of the simulated data sets, which for a sample of observations $\{x_1, \ldots, x_n\}$ is defined as follows: 
\begin{equation}\label{eqn-ecfm}
C(k)=\frac{1}{k}\sum_{i=1}^k(x_i-\bar{x})^4,~k=1,2,...,n,
\end{equation}
where $\bar{x}$ is the mean of the random sample. 

\begin{figure}[htb]
\centering \resizebox{0.46\textwidth}{!}{\includegraphics{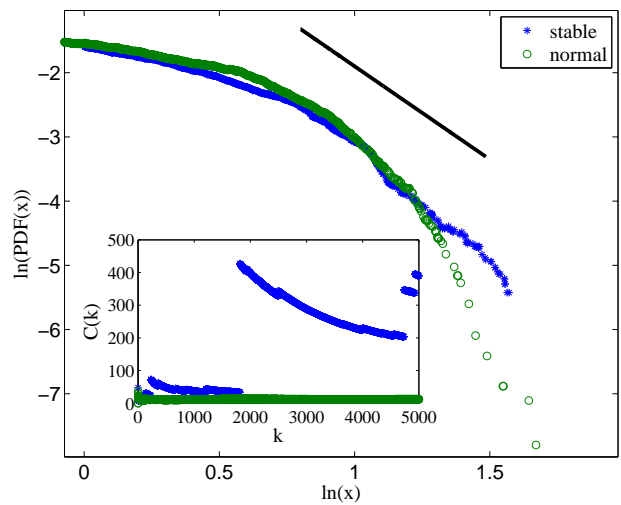}}
\caption{(Color online) Empirical PDF calculated for simulated sample from  L\'evy stable distribution with parameters $\alpha=1.9$, $\beta=0$, $\sigma=1$ and $\mu=0$ and for Gaussian distribution (with mean equal zero and variance equal $2$) in the log-log scale. Number of observations in both samples is $5000$. The straight line shows
the asymptotic slope of the L\'evy stable PDF with $\alpha=1.9$. Inset: Empirical cumulative fourth moment for the samples considered.}\label{alfa19}
\end{figure}

\begin{figure}[htb]
\centering \resizebox{0.46\textwidth}{!}{\includegraphics{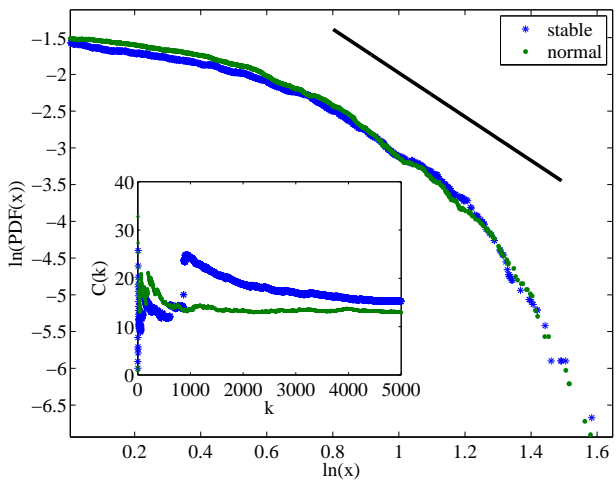}}
\caption{(Color online) Empirical PDF calculated for simulated sample from L\'evy stable distribution with parameters $\alpha=1.98$, $\beta=1$, $\sigma=1$ and $\mu=0$ and for  normal distribution (with mean equal zero and variance equal $2$) in the log-log scale. Number of observations in both samples is $5000$. The straight line shows
the asymptotic slope of the L\'evy stable PDF with $\alpha=1.9$. Inset: Empirical cumulative fourth moment for the samples considered. }\label{fig22}
\end{figure}

The formula (\ref{eqn-ecfm}) can be calculated for any sample obtained from an arbitrary probability distribution. For a fixed $k$ it forms a random variable. For distributions with finite fourth moment (e.g., Gaussian), the ECFM, as a function of $k$, converges to a constant, whereas for distributions with infinite fourth moment (e.g., L\'evy stable with $\alpha  < 2$) it diverges to infinity. The latter, for a finite sample, can be observed as an irregular chaotic behavior. Such reasoning stands for the idea of the first step of our algorithm. This simple test, which is presented in Fig. \ref{fig11} (inset), clearly indicates that there is a noticeable difference between L\'evy and Gaussian distributions. For the former the ECFM does not tend to a constant value with number $k$ of observations increasing, and behaves chaotically, while for the latter one for large numbers of observations it goes to the theoretical fourth moment that in this case is equal to $12$. Now, if we take stably distributed variables with smaller L\'evy index, say,
$\alpha = 1.9$, and the same total number of observations $n$, then the difference between the empirical L\'evy stable PDF and Gaussian PDF becomes more
visible; this is shown in Fig. \ref{alfa19}. Of course, in this case  we also observe the chaotic behavior
of the ECFM for
stably distributed variables, see the inset in Fig. \ref{alfa19}.  
In Fig. \ref{fig22} we demonstrate the right tail of the empirical PDF (in the log-log scale) for totally skewed L\'evy distribution with $\alpha = 1.98$, $\beta=1$, $\sigma=1$, and $\mu=0$. Number of observations is the same as in Figs. \ref{fig11} and \ref{alfa19}, i.e., $n = 5000$. Similarly to the symmetric case 
$\beta=0$, in Fig. \ref{fig22} we also plot the PDF of the Gaussian distribution with mean equal 
zero and variance equal to $2$, and the slope of the asymptotics of the L\'evy stable PDF. The difference between the considered PDFs is almost not visible, similar to Fig. \ref{fig11}. However, in contrast to Fig. \ref{fig11}, the visual test based on ECFM does not indicate clearly a difference between 
the analyzed distributions, 
since for both empirical PDFs the ECFM tends to a constant, see the inset in Fig. \ref{fig22}. This example 
demonstrates  that obviously there is a need to use a more advanced technique for recognition of the L\'evy stable distribution with $\alpha$ close to 2.

\section{Procedure of testing}\label{schematt}

In this section we present an algorithm how to recognize the difference between the Gaussian distribution and the
L\'evy stable distribution with the index $\alpha$ close to $2$. It is based on the empirical fourth moment studied in detail in Section \ref{simul}
and Anderson-Darling  and Jarque-Bera statistical tests.

Goodness-of-fit statistical tests, in general, are constructed to check whether a hypothetical distribution can be rejected for the given data or not. Their idea basically relies on checking how probable is reaching the value of some statistic calculated for the data if the underlying distribution was the hypothetical one.

Statistics usually measure the distance between the empirical and the fitted analytical cumulative distribution functions. For a sample of observations $\{x_1, \ldots, x_n\}$ the empirical cumulative distribution function  $F_n(x)$ is a piecewise constant function starting from zero with jumps of size $1/n$ at points $x_i$. The distance is usually measured either by a supremum or a quadratic norm \cite{dagste86}. 

The most well-know supremum statistic is the Kolmogorov-Smirnov (KS) statistic. It is just the supremum of the set of distances:
\begin{equation}\label{eqn:kologorov_stat}
D=\sup\limits_x\left|F_n(x)-F(x)\right|,
\end{equation}
where $F(x)$ is the analytical cumulative distribution function.

The Cramer-von Mises family of statistics incorporates the idea of the quadratic norm. It is defined by:
\begin{equation}\label{eqn:loss:cramervonmises} Q=n\int\limits_{-\infty}^{\infty}\left\{F_n(x)-F(x)\right\}^2\psi(x)\mathrm{d}F(x),
\end{equation}
where $\psi(x)$ is a suitable function which gives weights to the squared difference $\left\{F_n(x)-F(x)\right\}^2$. When $\psi(x)=[F(x)\left\{1-F(x)\right\}]^{-1}$, Eq. (\ref{eqn:loss:cramervonmises}) yields the Anderson-Darling (AD) statistic. 

The tests derived from the KS and AD statistics are called Kolmogorov-Smirnov and Anderson-Darling, respectively. It is well-known that the KS test exhibits poor sensitivity to deviations from the hypothetical distribution that occur in the tails, whereas the AD test is considered as one of the most powerful when the fitted distribution departs from the true distribution in this area \cite{dagste86}. Since the fit in the tails is of crucial importance in the stable case, AD will be our choice when testing for the non-Gaussian stable law.

For testing the Gaussianity we propose to use the standard Jarque-Bera (JB) test \cite{jarber87}, which is different from the KS and AD. It is a goodness of fit test  of whether sample data have the skewness and kurtosis matching a normal distribution. The JB statistic is defined as:
\begin{equation}
J = \frac{n}{6}\left(S^2+\frac{(K-3)^2}{4}\right),
\label{eqn:jb_stat}
\end{equation}
where $S$ and $K$ are the sample skewness and kurtosis respectively, namely
\begin{equation}
\label{eqn:S_stat}
S = \frac{1/n\sum_{i=1}^{n}(x_i-\bar{x})^3}{\left(\sqrt{1/n\sum_{i=1}^{n}(x_i-\bar{x})^2}\right)^3},
\end{equation}
\begin{equation}
\label{eqn:K_stat}
K = \frac{1/n\sum_{i=1}^{n}(x_i-\bar{x})^4}{\left(\sqrt{1/n\sum_{i=1}^{n}(x_i-\bar{x})^2}\right)^2}.
\end{equation}

Equations (\ref{eqn:jb_stat})-(\ref{eqn:K_stat}), similarly to Eq. (\ref{eqn-ecfm}), can be used for any sample from an arbitrary probability distribution. The value of the JB statistic given by Eq. (\ref{eqn:jb_stat}) forms a random variable which converges to zero if the underlying distribution has skewness zero and kurtosis 3 (e.g., Gaussian). Any deviation from zero skewness and kurtosis equal 3 increases the JB statistic. For distributions with infinite kurtosis (e.g., L\'evy stable with $\alpha  < 2$) it diverges to infinity. The test is quite standard and implemented in various numerical packages, like, e.g., R or Matlab.

To employ any of the tests, first, we need to estimate the parameters of the hypothetical distribution. In the Gaussian case the standard method is the maximum likelihood, whereas in the non-Gaussian stable case we suggest the use of the regression method estimates \cite{Koutrouvelis and Kogon,wer04}.

The general test of fit is structured as follows. The hypothesis is that a specific distribution is acceptable, whereas the alternative is that it is not. Small values of its test statistic $T$ are evidence in favor of the hypothesis, large ones indicate its falsity. To see how unlikely such a large outcome would be if the hypothesis is true, we calculate the $p$-value by: $p\mbox{-value} = P(T \ge t)$, where $t$ is the statistic value for a given sample. It is typical to reject the hypothesis when a small $p$-value is obtained, like, e.g., below $3\%$ or $5\%$. To calculate $p$-values for KS and AD tests we apply the procedure proposed in \cite{ros02} and described in detail in \cite{burjanwer11}. The $p$-values for the JB test were evaluated in Matlab using its standard procedure ``jbtest''.

Now, we can combine the AD and JB tests with the test based on ECFM, to introduce an algorithm for recognizing alpha-stable L\'evy distribution 
with L\'evy index close to 2.

In the first step of the algorithm,  we propose to observe the ECFM of the  sample. If the ECFM tends to a constant, we check for the Gaussian distribution by using the JB test. If its $p$-value exceeds the confidence level (usually $5\%$), then we can assume the underlying distribution of time series is Gaussian. In this case we estimate its parameters by using the standard maximum likelihood estimation method. If the JB test shows the data can not be modeled by a Gaussian distribution, then we test them for the L\'evy stable distribution. If the AD test gives the $p$-value that exceeds the confidence level, then we can assume the time series can be descibed by the alpha-stable distribution. In this case we estimate its parameters via the regression approach \cite{wer04}. 

\begin{figure}[htb]
 \resizebox{0.46\textwidth}{!}{\includegraphics{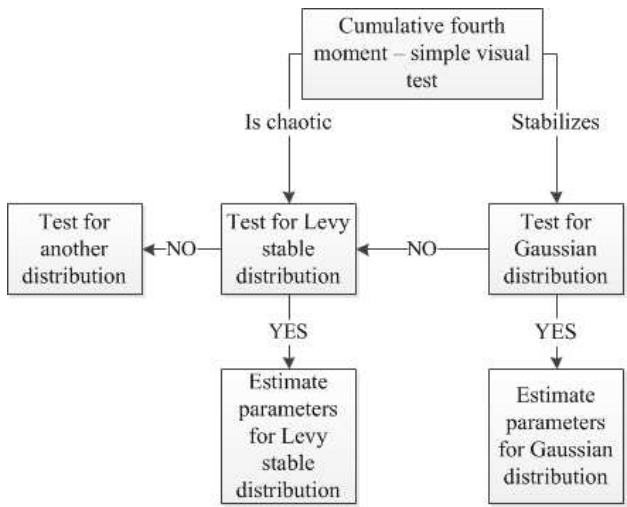}}
\caption{Schematic algorithm for recognition of L\'evy stable distribution with L\'evy index close to $2$. }\label{schemat}
\end{figure}

If the ECFM exhibits chaotic-like behavior, then we test for the L\'evy stable distribution. If the $p$-value is greater than the confidence level, then we can assume the data follow the  L\'evy stable law. Its parameters are regression method estimates. A scheme of the whole procedure is depicted in Fig. \ref{schemat}.

In order to illustrate the procedure we apply it to the simulated random samples from Section \ref{simul}. The obtained results are presented in Table \ref{tabsimul_data}. In the last two columns we demonstrate the values of the JB and AD statistics, respectively, together with the corresponding $p$-values (in parenthesis).
\begin{table}[tbp]
\begin{center}
\caption{Illustration of the introduced procedure for simulated samples from Section \ref{simul}. The last two columns contain values of the appropriate statistics, calculated by Eqs. (\ref{eqn:kologorov_stat})-(\ref{eqn:K_stat}), and the corresponding $p$-values (in parentheses). }
\label{tabsimul_data}
\begin{tabular}{|c|c|c|c|c|}
\hline\hline
\bf{Sample} &\bf{ECFM tends}&\bf{JB stat.}&\bf{AD stat. } \\
\bf{distribution} &\bf{to const? }&\bf{($p$-value)}&\bf{($p$-value)} \\
\hline
\mbox{L\'evy }&\mbox{NO}&-&0.24\\
\mbox{$\alpha=1.98$, $\beta=0$ }&&&(0.64)\\
\mbox{$\sigma=1$, $\mu=0$ }&&&\\\hline
\mbox{L\'evy }&\mbox{NO}&-&0.88\\
\mbox{$\alpha=1.90$, $\beta=0$ }&&&(0.42)\\
\mbox{$\sigma=1$, $\mu=0$ }&&&\\\hline
\mbox{L\'evy }&\mbox{YES}&61.33&0.18\\
\mbox{$\alpha=1.98$, $\beta=1$ }&&$(<0.01)$&(0.86)\\
\mbox{$\sigma=1$, $\mu=0$ }&&&\\
\hline
\mbox{Gaussian }&YES&0.55&-\\
\mbox{$\mu=0$, $\sigma=\sqrt{2}$ }&&(0.5)&\\ 
\hline \hline\end{tabular}
\end{center}
\end{table}

For the sample from the L\'evy stable distribution with parameters $\alpha=1.98$, $\beta=0$, $\sigma=1$ and $\mu=0$ we can observe chaotic behavior of the ECFM, see Fig. \ref{fig11}. Therefore, according to the procedure, we test its stability  using the AD test statistic. The obtained $p$-value indicates the time series constitutes independent identically distributed  (i.i.d.) L\'evy stable random variables. 

As we observe in Fig. \ref{alfa19}, the ECFM for the sample from symmetric L\'evy stable distribution with $\alpha=1.9$ and $\sigma=1$ behaves chaotically, therefore we use the AD statistic here. The obtained $p$-value clearly indicates the underlying stable distribution.

The ECFM for simulated data from the totally skewed stable distribution with $\alpha=1.98$ for large number of observations tends to a constant (see Fig. \ref{fig22}), therefore first we test the Gaussian law hypothesis. The JB test indicates the time series can not be modeled by the normal distribution, therefore we check for the L\'evy stable by using the AD test that justifies the stable distribution. 

For the last simulated sample from the normal distribution we observe that the ECFM tends to a constant. The obtained $p$-value clearly indicates that the underlying distribution of the analyzed time series is Gaussian.

The algorithm proposed in the paper provides a procedure to distinguish between the case  $\alpha\approx 2$ and $\alpha=2$. In order to distinguish the case of $\alpha\approx 2$  and  $\beta = 1$ with $\alpha\approx 2$  and $\beta  = 0$ one has to analyze the estimated skewness parameter $\beta$. However, we note that the confidence interval constructed for this parameter can be quite wide if the number of observations is not large enough. This circumstance makes the difference not definite.

\section{Applications}\label{appl}

\begin{figure}[htb]
\centering \resizebox{0.45\textwidth}{!}{\includegraphics{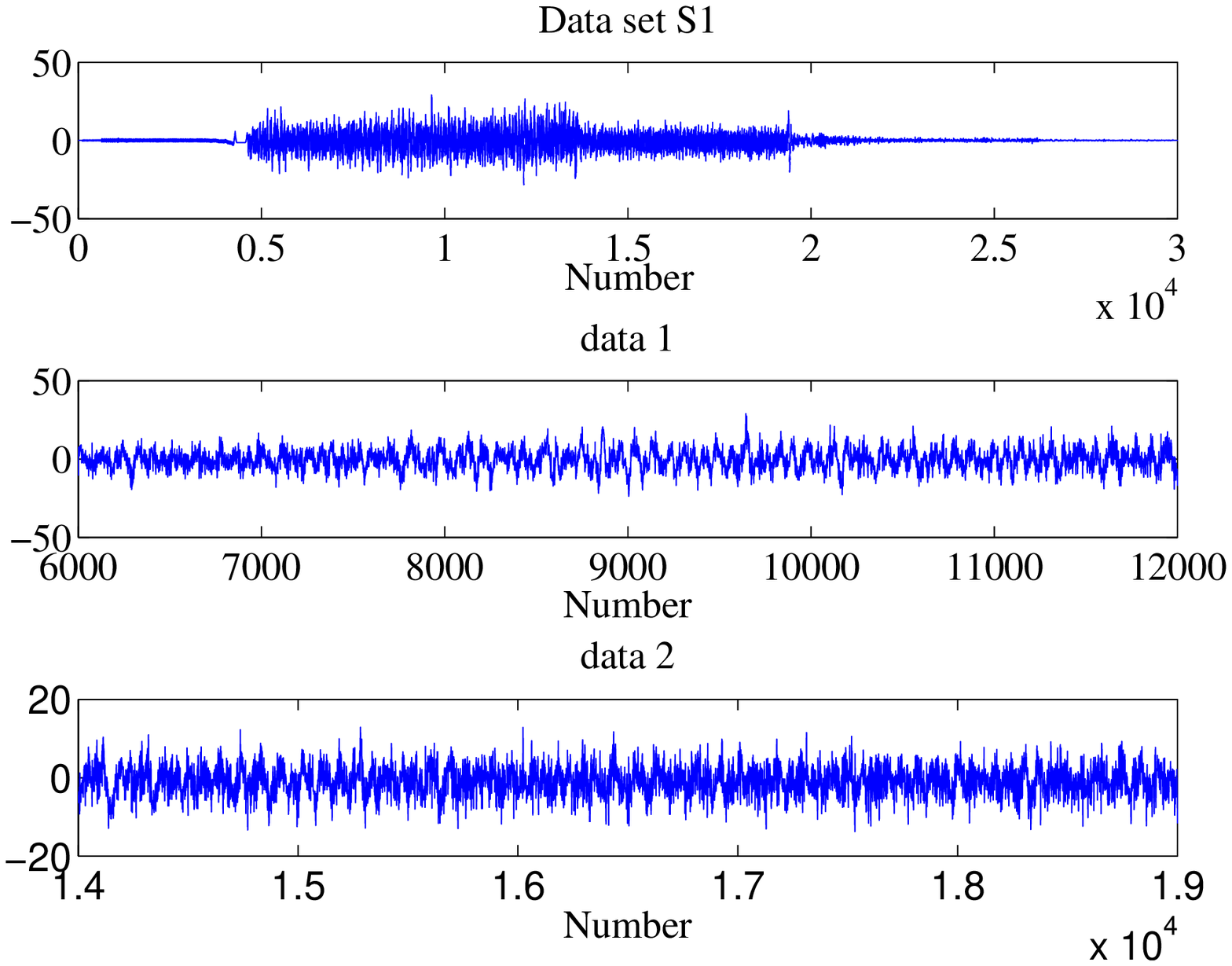}}
\centering \resizebox{0.45\textwidth}{!}{\includegraphics{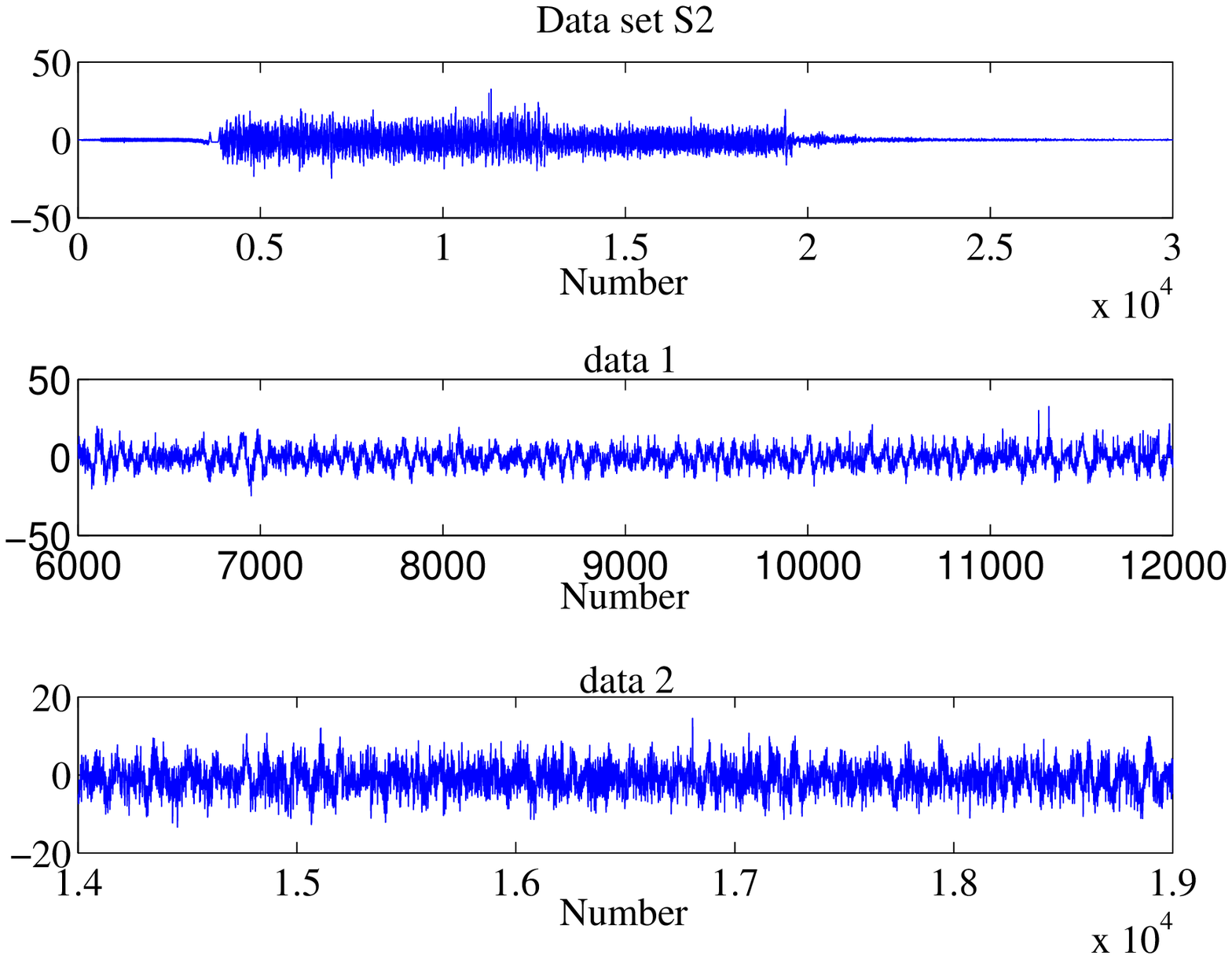}}
\centering \resizebox{0.45\textwidth}{!}{\includegraphics{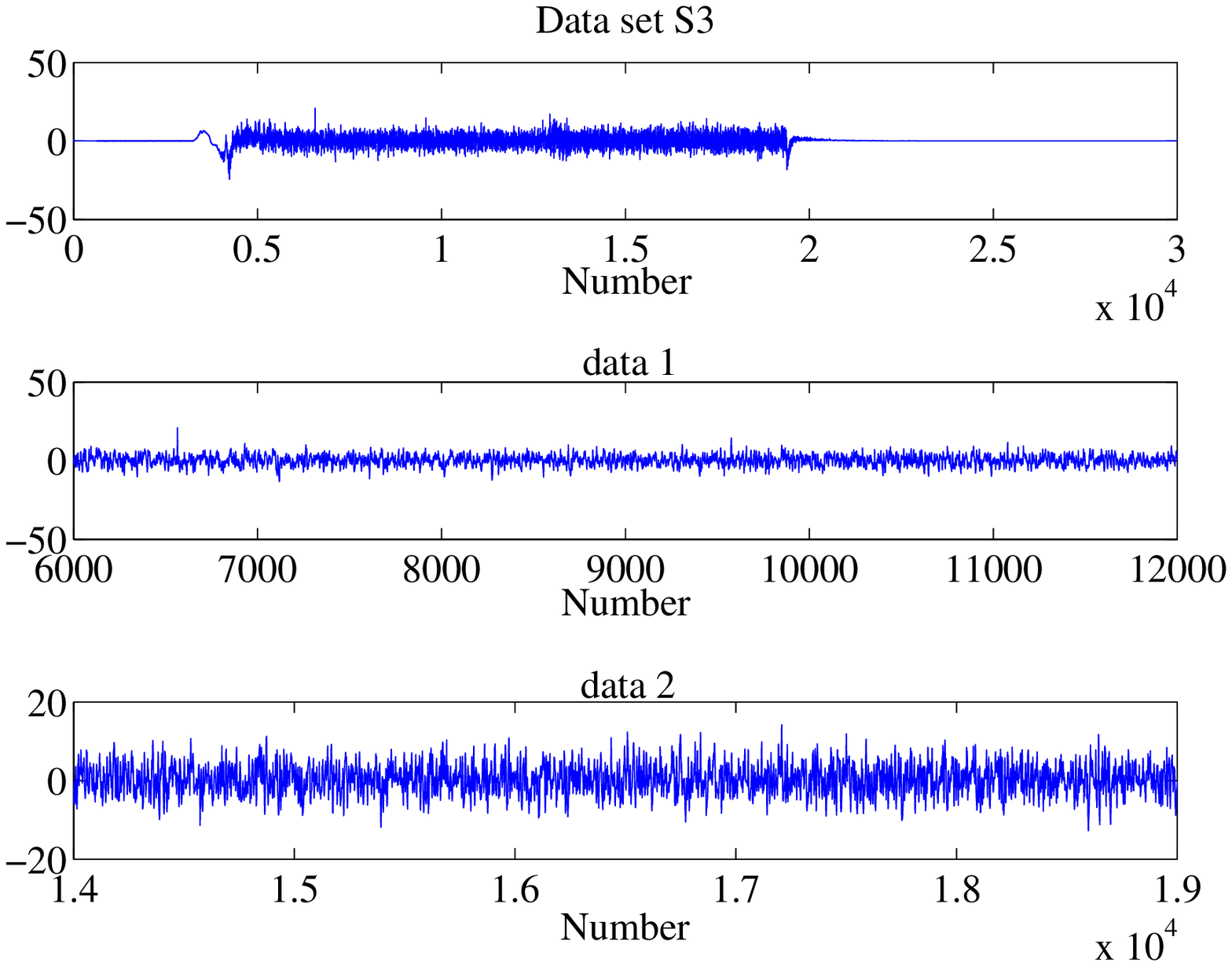}}
\caption{(Color online) The  empirical time series from plasma physics described in Section \ref{appl}. }\label{fig1_3}
\end{figure}

In this section we investigate the data obtained in experiment on controlled thermonuclear 
fusion device. Namely, we analyze the statistical properties of plasma fluctuations before 
and after so called L-H transition phenomenon, that is sudden transition from low confinement 
mode (L mode) to a high confinement mode (H mode) accompanied by suppression of turbulence 
and rapid drop of turbulent transport at the edge of thermonuclear device \cite{Wagner}. 
The implementation of the H mode regime, which is chosen as the operating mode for future 
ITER device \cite{iter}, requires detailed investigation of the transition physics. Electric 
Langmuir probes represent an effective diagnostic tool for this purpose \cite{probe}. The important characteristics of edge plasma turbulence, such as fluctuation amplitudes, spectra, and turbulence-induced transport are investigated in the Uragan-3M (U-3M) stellarator torsatron by the use of high resolution measurements of density (ion saturation current) and potential (floating potential) fluctuations 
with the help of movable Langmuir probe arrays \cite{Beletskii}. Changes in fluctuation behavior are good indicators of transition to improved confinement modes. In particular, investigations of spontaneous L–H-like transition (hereinafter, transition) in the U-3M torsatron reveal a distinct decrease of fluctuation level and radial turbulent particle flux \cite{VVChe}. In this section we analyze the three data sets. They present typical floating potential fluctuations (in volts) in turbulent plasma, registered in the U-3M torsatron for the small torus radial positions $r = 9.9$ cm (data set S1), $r = 10$ cm (S2), and $r = 11.25$ cm (S3). For the detailed description of the experimental set-up see Refs. \cite{Gonchar, Beletskii}. One can observe obvious change of the fluctuation amplitude in the middle position of every data source, which indicates the transition \cite{VVChe}. Therefore, we split each data set into two parts: before (data1) and after the transition point (data2). Moreover, we examine only such parts of each data which constitute the stationary process. Precisely, each data1 contains observations between 6000 and 12000, whereas data2 between 14000 and 19000. ADC sampling rate is 625 kHz that means the time resolution between two consecutive measurement points is 1.6 mcsec. In Fig. \ref{fig1_3} we depict the examined data sets and the selected parts before and after the L-H point.

\begin{figure}[thp]
\centering
 \resizebox{0.44\textwidth}{!}{\includegraphics{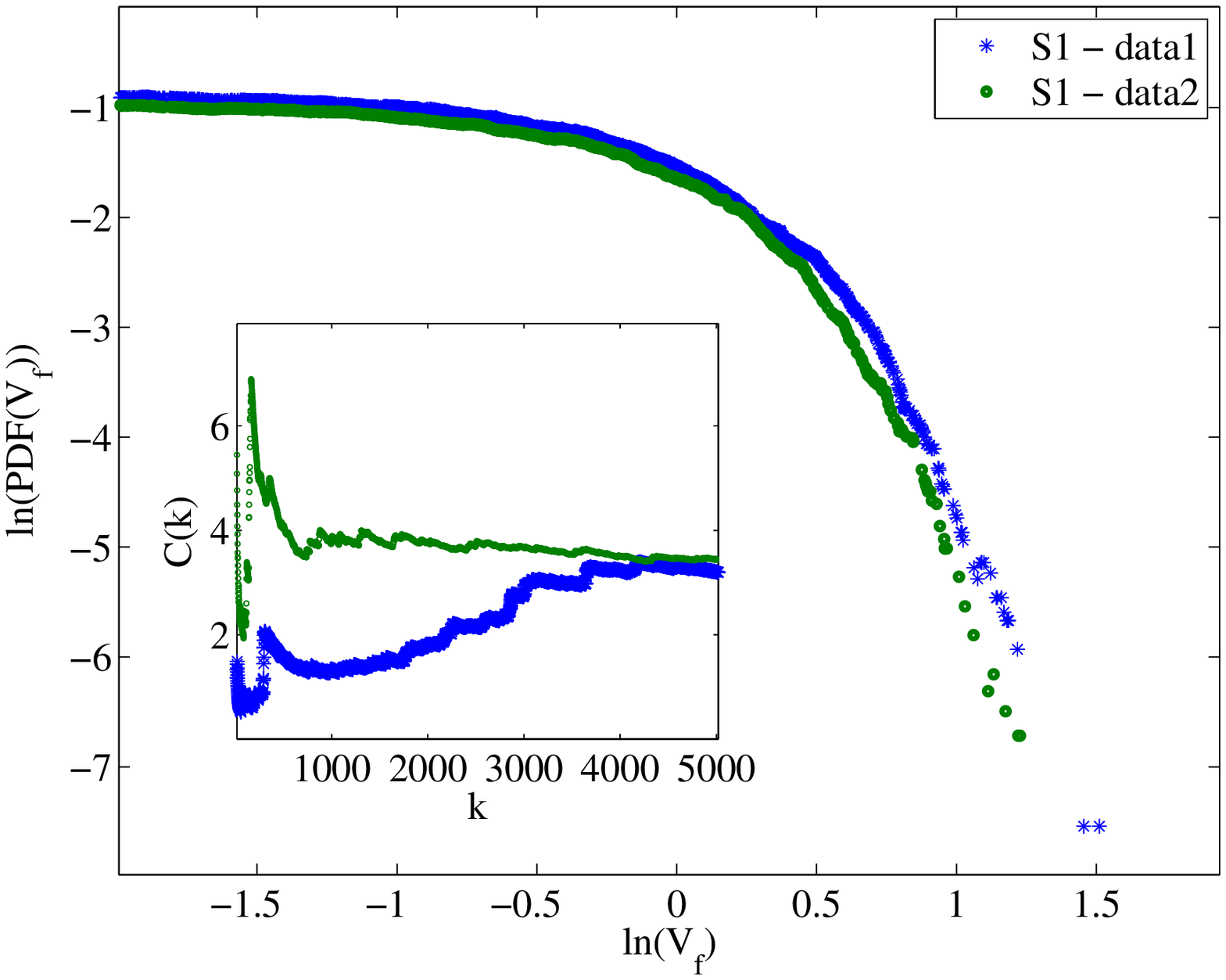}}
\centering
 \resizebox{0.44\textwidth}{!}{\includegraphics{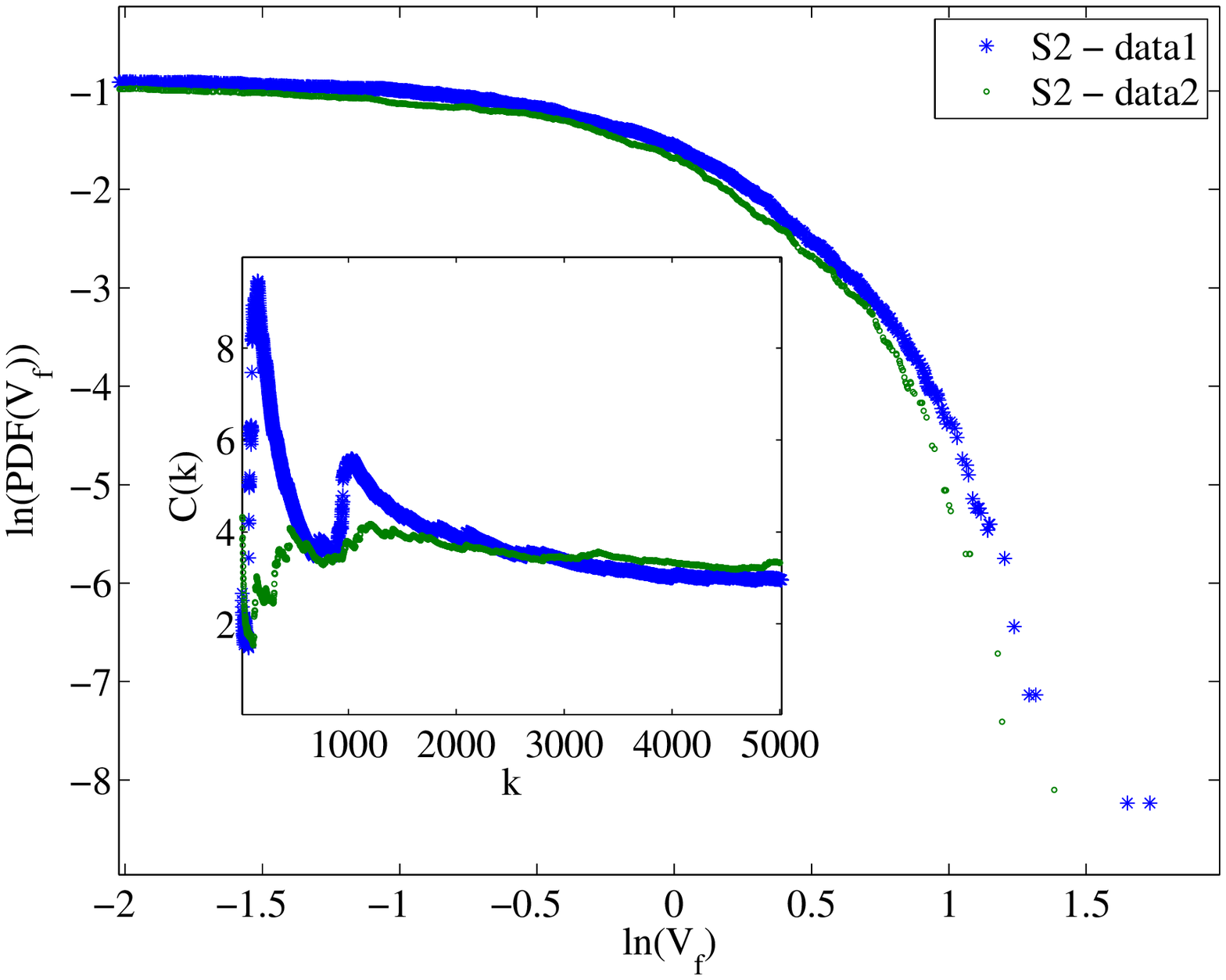}}
\centering
 \resizebox{0.44\textwidth}{!}{\includegraphics{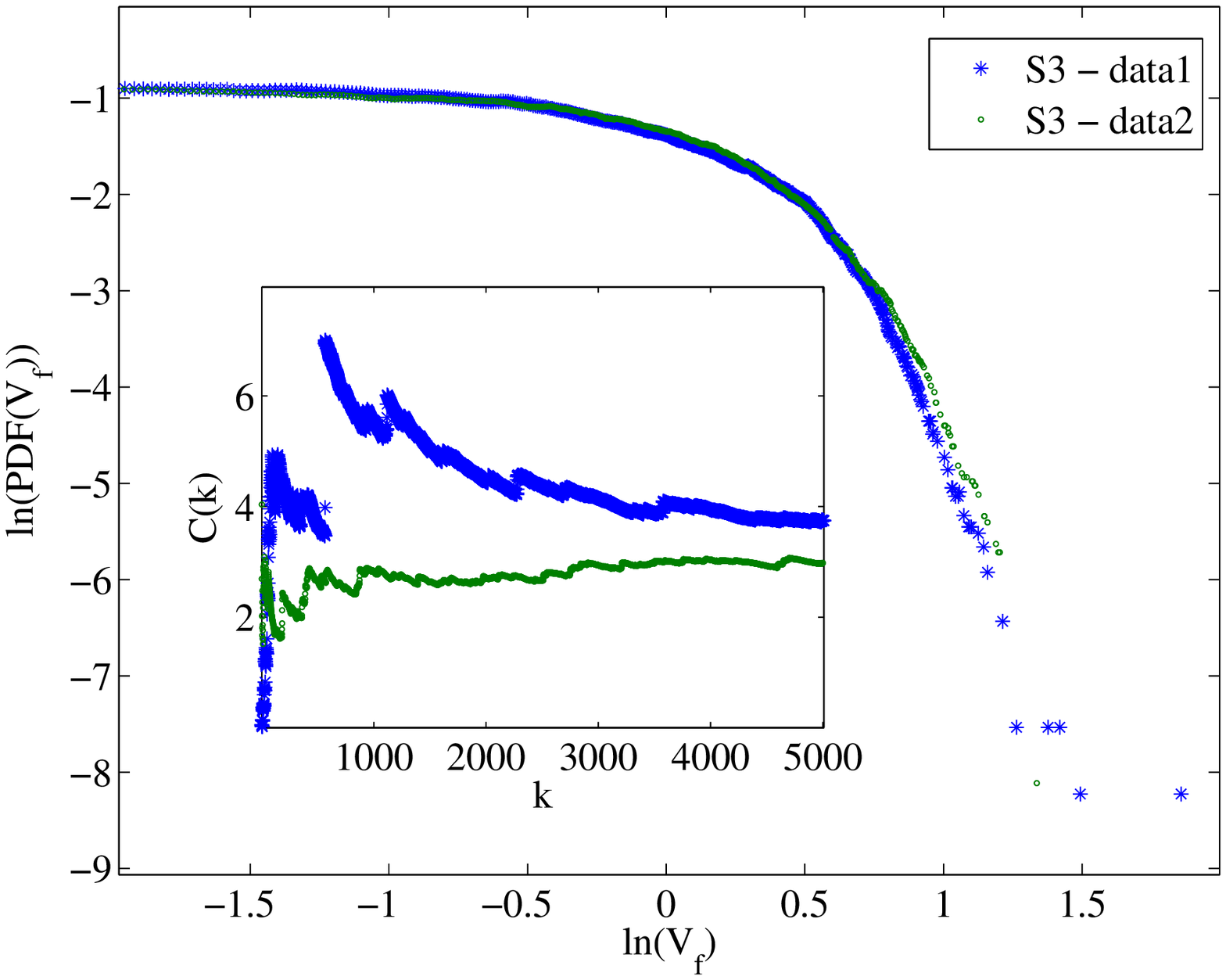}}
\caption{(Color online) The right tail of the empirical PDFs and ECFMs for six examined data sets.}
\label{mom3}
\end{figure}

According to the procedure of testing between L\'evy stable and Gaussian distributions presented in Section \ref{schematt}, we start with examining the ECFMs for all data sets. The result of this simple test together with right tails of the empirical PDFs are presented in Fig. \ref{mom3}.

This visual test indicates that following data sets can not be considered as a Gaussian time series: S1 -- data1, S2 -- data1 and S3 -- data1, whereas for the rest of the examined samples, i.e., S1 -- data2, S2 -- data2 and S3 -- data2, the ECFM stabilizes. In order to confirm this fact, we test all examined data sets for the Gaussian distribution by using the JB test presented in Section \ref{schematt}. The $p$-values and values of the JB statistics are depicted in Table \ref{tabb1}.
\begin{table}[tbp]
\begin{center}
\caption{Value of the test statistic and corresponding $p$-values (in parentheses) for the Gaussian law hypothesis  (the JB test).}\label{tabb1}
\begin{tabular}{|c|c|}
\hline\hline
\bf{Data set} &\bf{JB stat.}\\
 &\bf{($p$-value)}\\
\hline
\mbox{S1 -- data1}&$12.35$\\
&$(<0.01)$\\
\mbox{S1 -- data2}&$7.39$
\\&$(0.025)$\\\hline
\mbox{S2 -- data1}&$71.59$
\\&$ (<0.01)  $\\
\mbox{S2 -- data2}&$1.67$
\\&$(0.43)$\\\hline
\mbox{S3 -- data1}&$62.25$
\\&$  (<0.01)    $\\
\mbox{S3 -- data2}&$ 0.96 $
\\&$(0.5)$\\
\hline \hline
\end{tabular}
\end{center}
\end{table}

The JB test clearly indicates that S1 -- data1, S2 -- data1 and S3 -- data1 follow a non-Gaussian distribution because the $p$-values are smaller that the confidence level $5\%$ (in fact they are even less  than $1\%$). Let us now consider the sample S1 -- data2. In this case the ECFM  tends to a constant  whereas the JB test rejects the hypothesis of Gaussianity  with $p$-value close to the confidence level. This is why we use another test in order to confirm one of the hyphotheses (Gaussian or non-Gaussian distribution). We propose to use the AD test  presented in Section \ref{schematt}, for testing for Gaussianity. The value of the AD statistic is $0.5125$, while the corresponding $p$-value is equal to $0.16$, which indicates the Gaussian behavior. 

In the next step of our analysis we test 
the samples with chaotic ECFMs for the L\'evy distribution, namely S1 -- data1, S2 -- data1 and S3 -- data1.  In Table \ref{tab22} we present $p$-values and values of the AD statistics for the hypothesis of the stable distribution that indicate stable behavior of examined data sets.
\begin{table}[tbp]
\caption{Values of the test statistics and corresponding $p$-values (in parentheses) for the L\'evy stable law hypothesis (AD test).}
\begin{center}\label{tab22}
\begin{tabular}{|c|c|}
\hline\hline
\bf{Data set} &\bf{AD stat.}\\
\hline
\mbox{S1 -- data1}&$ 0.39$
\\&$(0.24)$\\
\hline
\mbox{S2 -- data1}&$0.48 $
\\&$(0.72)$\\
\hline
\mbox{S3 -- data1}&$0.62$
\\&$(0.91)$\\\hline\hline
\end{tabular}
\end{center}
\end{table}

Finally, we can present the index of stability and the skewness parameter for samples S1 -- data1, S2 -- data1 and S3 -- data1. They were obtained by the regression method \cite{Koutrouvelis and Kogon}. The estimated $\beta$ parameter for S1 -- data1 is equal to $0.2$ while for S2 -- data1  is on the level $1$ and for S3 -- data1 is $-1$. The results of the estimated parameter $\alpha$  and the corresponding confidence intervals are presented in Table \ref{tab33}. 
\begin{table}[tbp]
\caption{Estimated parameter $\alpha$ of the L\'evy stable distribution with the corresponding confidence intervals.}
\begin{center}\label{tab33}
\begin{tabular}{|c|c|}
\hline\hline
\bf{Data set} &$\alpha$\\
\hline
\mbox{S1 -- data1}&$1.98$\\&$[1.97   ; 1.99]$\\\hline
\mbox{S2 -- data1}&$1.97$\\&$[1.96   ; 1.99]$\\\hline
\mbox{S3 -- data1}&$1.98$\\&$[1.96  ;  1.99]$\\
\hline \hline
\end{tabular}
\end{center}
\end{table}

\section{Discussion}
\label{disc}

Let us first add a word of caution: of course, as with any other family of distributions
it is not possible to prove that a given data set is or is not stable \cite{Nolan99},
and even testing for normality is still an active field of research \cite{Brown}. In the
empirical data analysis like the one presented here the best what can be done is to determine 
whether or not the data are consistent with the hypothesis of stability. 

However, with the testing procedure developed in present paper we were able to distinguish between Gaussian and stable distribution with L\'evy index close to $2$.  In other words, we showed that the difference between such laws is essential and one can identify the alpha-stable distribution with $\alpha$ close to $2$, even if the number of available data points is not too large. We illustrated the procedure on simulated samples from Gaussian and non-Gaussian stable distributions. The results show that the algorithm is  reliable and works well.  It might be useful for detecting L\'evy stable distributions in different experiments.

As an example, we applied the developed procedure to assess the stable properties of the fluctuations
measured in the edge plasma of stellarator torsatron ``Uragan 3M'' and demonstrated that the 
statistics of fluctuations changes from stable to normal at the LH transition.

Another remark concerns the observed phenomenon of changing statistics. 
We first note that the L\'evy statistics of plasma fluctuations has been detected
in the measurements on stellarators ``Uragan 3M'' and Heliotron J by using the method of quantiles
\cite{Gonchar} and in the tokamak ADITYA by measuring probability to stay at the origin \cite{Kaw}.
The L\'evy indexes of the detected stable distributions were sufficiently less than
2. Non-Gaussian heavy tailed PDFs for low frequency turbulence have been also observed
in other toroidal plasma confinement systems such as T-10 tokamak, L-2M, TJ II and
LHD stellarators \cite{Skvortsova} as well as in astrophysical plasmas \cite{Chapman, Watkins}.
As to the physics of the change of the statistics, it in turn, implies qualitative 
changes of the basic properties of plasma turbulence, and should be taken into account in the
theoretical models of the LH transition. This important issue is out of the scope of the 
present paper. On the other hand, we believe that the change of statistics that we observed 
in the data from ``Uragan 3M'' may
inspire other authors to check whether LH transition is accompanied by change of statistics
in other plasma confinement devices as well. 

\begin{acknowledgments}
The authors thank V. Chechkin and L. Grigor'eva for providing us with the experimental data
and for discussing the results.
\end{acknowledgments}

\end{document}